\documentclass[12pt]{article}

\usepackage{amsmath}

\def\bdelta{\mbox{\boldmath $\delta$}}
\newcommand{\sfrac}[2]{{\textstyle{#1\over#2}}}

\title{\sc Simple expressions for second order density perturbations in standard cosmology}

\begin{document}

\author{ \\
{\Large\sc Claes Uggla}\thanks{Electronic address:
{\tt claes.uggla@kau.se}} \\[1ex]
Department of Physics, \\
University of Karlstad, S-651 88 Karlstad, Sweden
\and \\
{\Large\sc John Wainwright}\thanks{Electronic address:
{\tt jwainwri@uwaterloo.ca}} \\[1ex]
Department of Applied Mathematics, \\
University of Waterloo,Waterloo, ON, N2L 3G1, Canada \\[2ex] }

\date{}
\maketitle

\begin{abstract}

In this paper we present four simple expressions for the relativistic first
and second order fractional density perturbations for $\Lambda$CDM
cosmologies in different gauges: the Poisson, uniform curvature, total matter
and synchronous gauges. A distinctive feature of our approach is the use of a
canonical set of quadratic differential expressions involving an arbitrary
spatial function, the so-called comoving curvature perturbation, to describe
the spatial dependence, which enables us to unify, simplify and extend
previous seemingly disparate results. The simple structure of the expressions
makes the evolution of the density perturbations completely transparent and
clearly displays the effect of the cosmological constant on the dynamics,
namely that it stabilizes the perturbations. We expect that the results will
be useful in applications, for example, studying the effects of primordial
non-Gaussianity on the large scale structure of the universe.


\end{abstract}

\centerline{\bigskip\noindent PACS numbers: 04.20.-q, 98.80.-k, 98.80.Bp,
98.80.Jk}

\section{Introduction}

Increasingly accurate observations of the cosmic microwave background (CMB)
and the large scale structure (LSS) of the universe are becoming available
and are influencing developments in cosmology. Analyzing these observations
necessitates the theoretical study of possible deviations from linearity, for
example, how primordial non-Gaussianity affects the anisotropy of the CMB and
the LSS, using nonlinear perturbations of Friedmann-Lemaitre-Robertson-Walker
(FLRW) cosmologies (see, e.g~\cite{baretal04b, baretal10b, pitetal10}). Much
of the theoretical work has dealt with nonlinear perturbations of flat FLRW
cosmologies with dust (a matter-dominated universe, see
e.g~\cite{matetal98,baretal05}) and more recently, also with a cosmological
constant ($\Lambda$CDM cosmologies, see
e.g~\cite{tom05,baretal06,baretal10,bruetal13}). One aspect of the work is to
provide an expression for the second order fractional density perturbation
${}^{(2)}\!{\bdelta}$ that can be used for comparison with observations, and
this is the focus of the present paper.

In a previous paper~\cite{uggwai13} we gave a new expression for
${}^{(2)}\!{\bdelta}$ in the Poisson gauge, also including the effects of
spatial curvature (see equations (33)-(36) in~\cite{uggwai13}). The evolution
of ${}^{(2)}\!{\bdelta}$ is determined by four functions of time that are
integrals involving a Green's function, and these are the main source of the
complexity of the expression. Subsequently we realized that if the background
spatial curvature is zero (which we will assume in this paper), these
functions can be simplified significantly, leading to a physically
transparent form of the expression in the Poisson gauge. Then we derived the
expressions for ${}^{(2)}\!{\bdelta}$ in other gauges by using gauge
transformation rules. We hope that our final simple expressions will be
useful for future work relating LSS observations to theory. The goal of this
paper is to give a concise overview of these results in an easily accessible
form, leaving the derivations to a separate paper~\cite{uggwai13c}, and to
use these results to perform an asymptotic analysis. We have made the present
paper essentially self-contained, but refer the reader to~\cite{uggwai13c}
for full details about the background. Throughout we have made two
simplifying restrictions: first that \emph{the perturbations at the linear
order are purely scalar}, and second that \emph{the decaying mode is zero}.

The outline of the paper is as follows. In the next section we give a unified
expression for the  first order fractional density perturbation in the four
gauges and in section~\ref{Sec:secgen} we give the corresponding second order
expressions, including the special case when the background cosmology is
Einstein-de Sitter. Section~\ref{Sec:concl} contains a discussion of the
results, focussing on the dynamics of the perturbations and the effect of the
cosmological constant, and  the gauge-robustness of the results. A comparison
with related work in the literature is also given.

\section{First order perturbations}

The background cosmology is a flat FLRW universe with scale factor $a$ and
Hubble scalar $H$, containing dust with background matter density
${}^{(0)}\!\rho_m$ and a cosmological constant $\Lambda$. The matter-energy
content is described by the usual density parameters:
\begin{equation} \Omega_m=\frac{^{(0)}\!\rho_m}{3H^2}, \qquad
\Omega_{\Lambda}=\frac{\Lambda}{3H^2}, \end{equation}
which satisfy $\Omega_m + \Omega_{\Lambda}=1$.\footnote{We use units such
that $8\pi G = 1 = c$.} We use the dimensionless scale factor $x:=a/a_0$,
normalized at some reference epoch $t_0$, as time variable (when $t_0$ is the
present time, $x = (1+z)^{-1}$, where $z$ is the redshift). The conservation
law shows that $a^3\,{}^{(0)}\!\rho_m$ is constant, which we write in terms
of $\Omega_m$ and the dimensionless Hubble scalar ${\cal H}:=aH$ as ${\cal
H}^2 x\,\Omega_m=m^2$. Setting $x=1$ shows that the constant $m$ is given by
$m^2={\cal H}_0^2 \,\Omega_{m,0}$. It follows that
\begin{equation}  {\cal H}^2={\cal H}^2_0\,( \Omega_{\Lambda,0}\,x^2 + \Omega_{m,0}\, x^{-1}). \label{calH} \end{equation}
We will use $x$, ${\cal H}$, $\Omega_m$ and the constant $m$ to describe the
influence of the background geometry on the perturbations.

We now describe the \emph{linear perturbations}, subject to the above
restrictions. In the Poisson gauge the Bardeen potential is given by:
\begin{equation}\label{psi_p}
{}^{(1)}\!{\Psi}_{\mathrm p}(x,x^i) = g(x)\zeta(x^i),
\end{equation}
where\footnote{See~\cite{uggwai13c}, equation (8). The function $g(x)$, up to
a constant multiple, is referred to as the growth suppression factor. See for
example~\cite{baretal10}, following equation (11).}
\begin{equation}\label{fg}
g(x):= \sfrac32 m^2
\frac{{\mathcal H}}{x^2}\int_0^x \frac{d{\bar x}}{{\mathcal H}({\bar x})^3},
\end{equation}
and $\zeta(x^i)$ is the linear comoving curvature perturbation, which is
exactly conserved for $\Lambda$CDM cosmologies.\footnote{See~\cite{uggwai12},
page 15. $\zeta$ is sometimes denoted by ${\cal R}$, see for example eq.
(7.46) in~\cite{malwan09}.} The fractional density perturbation
${}^{(1)}\!{\bdelta}_{\bullet}$, where the bullet identifies the gauge, has
the following common structure in the Poisson, uniform curvature, total
matter and synchronous gauges:
\begin{subequations}\label{bdeltaall}
\begin{equation} {}^{(1)}\!{\bdelta}_{\bullet} = A_{\bullet}\zeta + \sfrac23m^{-2} xg{\bf D}^2
\zeta,  \label{bdelta_1} \end{equation}
where ${\bf D}^2$ is the spatial Laplacian, defined in eq.~\eqref{2order_D},
while the expressions for $A_{\bullet}$ in the different gauges are as follows:
\begin{equation} A_{\mathrm p} =-3(1-g), \qquad A_{\mathrm c} =-3,   \qquad
A_{\mathrm v} =A_{\mathrm s} = 0,  \label{bdel_v1} \end{equation}
\end{subequations}
where $_{\mathrm p}$, $_{\mathrm c}$, $_{\mathrm v}$ and $_{\mathrm s}$
indicate the Poisson, uniform curvature, total matter and synchronous gauges,
respectively ((see~\cite{uggwai13c}, equation (12)).

\section{Second order perturbations}  \label{Sec:secgen}

We now come to \emph{second order perturbations}. The second order Bardeen
potential in the Poisson gauge is given by\footnote{See equation (23a)
in~\cite{uggwai13c}, specialized to the case of a flat background as in
section 4.3 in that paper.}
\begin{equation}  {}^{(2)}\!\Psi_{\mathrm p}= g\!\left({\cal B}_1\zeta ^2+ {\cal B}_2 {\cal
D}(\zeta)+\sfrac23 m^{-2}xg\left[{\mathcal B}({\bf D}\zeta)^2 + (1-{\mathcal
B}){\bf D}^2 {\cal D}(\zeta)\right]\right),     \label{psi_2}  \end{equation}
where ${\cal B}_1, {\cal B}_2$ and ${\cal B}$ are functions of $x$, and $g$
is the first order perturbation function~\eqref{fg}. The spatial dependence
is specified by four quadratic differential expressions involving the
comoving curvature perturbation $\zeta$, where
\begin{equation}\label{DA}
\left({\bf D}{\zeta}\right)^2:=\gamma^{ij}({\bf D}_i {\zeta})( {\bf D}_j {\zeta}), \qquad
{\cal D}({\zeta}):=\sfrac32 {\bf D}^{-4}{\bf D}^{ij} ({\bf D}_ i {\zeta})({\bf D}_j {\zeta}).
\end{equation}
Here ${\bf D}_i$ denotes covariant differentiation with respect to the
background spatial metric $\gamma_{ij}$\footnote{In the present case
$\gamma_{ij}$ is the Euclidian metric, which in Cartesian coordinates yields
${\bf D}_i = \partial_i$. However, since~\cite{uggwai13} and~\cite{uggwai13c}
also deal with non-zero background spatial curvature, we keep the notation of
those papers.} and ${\bf D}^{-2}$ denotes the inverse Laplacian. The
Laplacian and the symmetrized trace-free second derivative operator are
defined by
\begin{equation} \label{2order_D}
{\bf D}^2 := \gamma^{ij}{\bf D}_i{\bf D}_j, \qquad {\bf
D}_{ij} := {\bf D}_{(i}{\bf D}_{j)} - \sfrac13 \gamma_{ij}{\bf D}^2.
\end{equation}
The functions ${\cal B}_1$ and ${\cal B}_2$  are algebraic expressions in
terms of $g(x)$:\footnote{The expression~\eqref{psi_2} for
${}^{(2)}\!\Psi_{\mathrm p}$ includes the initial conditions and is given by
equations (23a) and (38) in~\cite{uggwai13c}, with $C_{\mathrm{nl}}$ given by
(42). Thus the functions~\eqref{B_final} are obtained from ${\cal B}_1(x)$
and ${\cal B}_1(x)$ in~\cite{uggwai13c} when $K=0$, as given by equations
(45c,d), by adding $\sfrac45- 2 a_{\mathrm{nl}}$ and $\sfrac85$,
respectively.}
\begin{subequations} \label{B_final}
\begin{align}
{\cal B}_1(x) &= 1 - 2 a_{\mathrm{nl}} - g +\sfrac32\Omega_mg^{-1} (1-g)^2, \label{B1}  \\
{\cal B}_2(x) &= 4(1- g), \label{B2}
\end{align}
\end{subequations}
while ${\mathcal B}$ is an integral involving $g(x)$ and the
background variables:
\begin{equation} {\mathcal B}(x)=\frac{{\cal H}(x)}{2x^3g(x)^2}\int_0^x
\frac{{\bar x}^2\,\Omega_m}{\cal H}\left( g^2-\sfrac32\Omega_m (1-g)^2  \right)d{\bar x}, \label{cal_Bfinal}
\end{equation}
where $g, \Omega_m$ and ${\cal H}$ inside the integral are functions of
${\bar x}$ (see equation (52) in~\cite{uggwai13c}).

The constant $a_{\mathrm{nl}}$ in~\eqref{B1} describes the level of
primordial non-Gaussianity on super-horizon scales. It enters into the
solution of the second order perturbation equations through the values of the
metric and matter perturbations at the end of inflation which act as initial
conditions by restricting the first and second order conserved
quantities\footnote{See~\cite{malwan09}, equations (7.61) and (7.71). The
subscript $_\rho$ stands for the uniform density gauge, and the subscript
$_\mathrm {mw}$ stands for Malik-Wands.}
${}^{(1)}\!\zeta_{\mathrm{mw}}:=-{}^{(1)}\!\Psi_{\rho}$ and
${}^{(2)}\!\zeta_{\mathrm{mw}}:=-{}^{(2)}\!\Psi_{\rho}$ according to
%
\begin{equation}
{}^{(2)}\!\zeta_{\mathrm{mw}}=2a_{\mathrm{nl}}\left({}^{(1)}\!\zeta_{\mathrm{mw}}\right)^2, \label{png}
\end{equation}
on super-horizon scales. The second order perturbation equations determine
${}^{(2)}\!\Psi_{\mathrm p}$ up to an arbitrary spatial function $C(x^i)$,
which is determined by~\eqref{png}, thereby introducing $a_{\mathrm{nl}}$
into the solution. The absence of primordial non-Gaussianity corresponds to
$a_{\mathrm{nl}}=1$. We refer to section 3.5 in~\cite{uggwai13c} and other
references given there for details.

The functions $g(x)$ and ${\cal B}(x)$, which determine the time dependence
of ${}^{(2)}\!\Psi_{\mathrm p}$, also determine the time dependence of the
second order fractional density perturbation ${}^{(2)}\!{\bdelta}_{\bullet}$.
In order to give a unified description of the spatial dependence of
${}^{(2)}\!{\bdelta}_{\bullet}$ we introduce the following eight `canonical'
quadratic differential expressions:
\begin{equation}  \zeta^2, \quad {\cal D}(\zeta), \quad ({\bf D}\zeta)^2, \quad {\bf
D}^2{\cal D}(\zeta), \quad {\bf D}^2 \zeta^2, \quad {\bf D}^2({\bf
D}\zeta)^2, \quad {\bf D}^4{\cal D}(\zeta), \quad ({\bf D}^2\zeta)^2,
\label{quad_zeta}   \end{equation}
the first four of which appear in ${}^{(2)}\!\Psi_{\mathrm p}$. The
first seven of these determine the spatial dependence of
${}^{(2)}\!{\bdelta}_{\bullet}$ in the Poisson, uniform curvature and total matter
gauges:
\begin{equation} \begin{split} {}^{(2)}\!{\bdelta}_{\bullet} &= \quad \underbrace{A_{1,{\bullet}}\zeta^2 +
A_{2,{\bullet}}{\cal D}(\zeta)}_{\text{the super-horizon part}} \\
&\quad  + \underbrace{\sfrac23 m^{-2}xg\left[A_{3,{\bullet}}({\bf D}\zeta)^2 +
A_{4,{\bullet}}{\bf
D}^2{\cal D}(\zeta) +A_{5,{\bullet}}{\bf D}^2 \zeta^2 \right]}_{\text{the post-Newtonian part}}  \\
&\quad  + \underbrace{\sfrac49m^{-4}x^2 g^2\left[{\mathcal B}\,{\bf D}^2({\bf D}\zeta)^2 +
(1-{\mathcal B}){\bf D}^4{\cal D}(\zeta)  \right]}_{\text{the Newtonian part}},
\end{split}  \label{delta_2}  \end{equation}
where the bullet indicates the gauge. The time dependence is
contained in the coefficients ${\mathcal B}$ and $A_{i,{\bullet}}$, $i=1,\dots,5$,
which are functions of the scale factor $x$. The three labeled parts
in~\eqref{delta_2} are identified by the weight of the operator ${\bf D}_i$
acting on $\zeta(x^i)$. For the \emph{super-horizon part}, ${\bf D}_i$ has
weight zero, for the \emph{post-Newtonian part} it has weight 2, and for the
\emph{Newtonian part} it has weight 4.

We obtain the coefficients $A_{i,{\bullet}}$ in~\eqref{delta_2} by setting
the background spatial curvature to zero ($K=0$) in the general expressions
in~\cite{uggwai13c} (see section 4.3 in that paper for details). In the
\emph{Poisson gauge} we have
\begin{subequations}  \label{Ai_p}
\begin{align}
A_{1,{\mathrm p}} &= 3(1-g)\left( 1 +2a_{\mathrm{nl}} - 4g +\sfrac32\Omega_m(1-g) \right), \\
A_{2,{\mathrm p}} &= 12g(1-g),  \\
A_{3,{\mathrm p}} &=3g({\mathcal B}-2)-\sfrac32\Omega_m\, g^{-1}(1-g)^2 , \\
A_{4,{\mathrm p}} &= -2 -3g({\mathcal B}-1) -\sfrac32\Omega_m\,g^{-1}(1-g)^2 , \\
A_{5,{\mathrm p}} &=1-2a_{\mathrm{nl}} + 3g+ \sfrac32\Omega_m\,g^{-1}(1-g)^2,
\end{align}
\end{subequations}
in the \emph{uniform curvature gauge},
\begin{subequations}  \label{Ai_c}
\begin{align}
A_{1,{\mathrm c}} &=3(1+2 a_{\mathrm{nl}}), \\
A_{2,{\mathrm c}} &=0, \\
A_{3,{\mathrm c}} &=- \sfrac32 \Omega_m\,g^{-1}, \\
A_{4,{\mathrm c}} &= - 2 -\sfrac32 \Omega_m\,g^{-1}(1-2g), \\
A_{5,{\mathrm c}} &= 1-2 a_{\mathrm{nl}} +\sfrac32 \Omega_m\,g^{-1}(1-g),
\end{align}
\end{subequations}
and in the \emph{total matter gauge},
\begin{equation} (A_1, A_2, A_3, A_4,A_5)_{\mathrm v}= (0, 0, -5, 0,
2(2-a_{\mathrm{nl}})). \label{Ai_v}   \end{equation}
For the \emph{synchronous gauge} we have (equation (55) in~\cite{uggwai13c})
\begin{equation} \begin{split}
{}^{(2)}\!{\bdelta}_{\mathrm s}&=
\sfrac23 m^{-2}xg\left[-5({\bf D}\zeta)^2  + 2(2-a_{\mathrm{nl}}) {\bf D}^2 \zeta^2 \right] \\
&\qquad + \sfrac49m^{-4}x^2g^2\left[({\mathcal B}+\sfrac13)\left({\bf D}^2({\bf D}\zeta)^2  -
{\bf D}^4{\cal D}(\zeta)\right) +2({\bf D}^2\zeta)^2 \right].
\end{split}
\label{deltaN_2s}
\end{equation}
Note the additional term $({\bf D}^2\zeta)^2$ in the Newtonian part, which
modifies the form~\eqref{delta_2}.

\subsection{Einstein-de Sitter as background}

In the limiting case $\Lambda=0$ the background is the Einstein-de Sitter
universe, which satisfies $\Omega_m=1, \Omega_{\Lambda}=0$ and ${\cal H}^2
x=m^2$. It follows from~\eqref{fg} and~\eqref{cal_Bfinal} that the key
functions $g(x)$ and ${\mathcal B}(x)$ simplify to
\begin{equation}  g(x)=\sfrac35,\quad {\mathcal B}(x)=\sfrac{1}{21}. \label{Eds_3} \end{equation}
Substituting~\eqref{Eds_3} into~\eqref{Ai_p},~\eqref{Ai_c} and~\eqref{Ai_v}
shows that ${}^{(2)}\!{\bdelta}_{\bullet}$ has the following form for the
Poisson, uniform curvature and total matter gauges:
\begin{equation}\begin{split} {}^{(2)}\!{\bdelta}_{\bullet} &= A_{1,{\bullet}} \zeta^2+ A_{2,{\bullet}}{\cal D}(\zeta)\\
& \quad +\sfrac25m^{-2}x\left[A_{3,{\bullet}}({\bf D}\zeta)^2 +
A_{4,{\bullet}}{\bf D}^2{\cal D}(\zeta) + A_{5,{\bullet}}{\bf D}^2 \zeta^2 \right]  \\
& \quad + \sfrac{4}{25}m^{-4}x^2\left[\sfrac{1}{21}{\bf D}^2({\bf D}\zeta)^2
+ \sfrac{20}{21}{\bf D}^4{\cal D}(\zeta)  \right],
\end{split}  \label{delta_2_eds}  \end{equation}
where the coefficients $A_{i,{\bullet}}$ specialize to
\begin{subequations}
\begin{align}
({ A}_1,{A}_2,{ A}_3,{ A}_4,{ A}_5)_{\mathrm p}&=
(-\sfrac{12}{5}(\sfrac25 - a_{\mathrm{nl}}),\sfrac{72}{25},-\sfrac{137}{35},-\sfrac{24}{35},2(\sfrac85 - a_{\mathrm{nl}})),   \label{eds_p} \\
({ A}_1,{A}_2,{ A}_3,{ A}_4,{ A}_5)_{\mathrm c}&= (3(1+2a_{\mathrm{nl}}), 0,-\sfrac52,-\sfrac{3}{2},2(1- a_{\mathrm{nl}})),   \label{eds_c}  \\
({ A}_1,{A}_2,{ A}_3,{ A}_4,{ A}_5)_{\mathrm v}&= (0,0,-5,0,2(2- a_{\mathrm{nl}})).  \label{eds_v}
\end{align}
\end{subequations}
For the synchronous gauge, using~\eqref{deltaN_2s} and~\eqref{Eds_3}, we
obtain
\begin{equation} \begin{split} {}^{(2)}\!{\bdelta}_{\mathrm s} &=  \sfrac25m^{-2}x\left[-5({\bf D}\zeta)^2
+  2(2-a_{\mathrm{nl}}){\bf D}^2 \zeta^2 \right]  \\
&\quad  +\sfrac{4}{25}m^{-4} x^2\left[\sfrac{8}{21} \left({\bf D}^2({\bf
D}\zeta)^2  - {\bf D}^4{\cal D}(\zeta)\right) + 2({\bf D}^2\zeta)^2 \right].
\end{split}  \label{delta_2s_eds}  \end{equation}

This case has a lengthy history, starting with Tomita~\cite{tom67}.
Expressions for ${}^{(2)}\!{\bdelta}_{\bullet}$ in the synchronous gauge have
been given in~\cite{tom67} (eq. (5.1)),~\cite{matetal98} (eq. (4.33))
and~\cite{baretal05} (eq. (7)), and in the Poisson gauge in~\cite{matetal98}
(eq. (6.10)),~\cite{baretal05} (eq. (8)) and~\cite{hwaetal12} (eq. (43)). The
different expressions are not related in an obvious way. However, after
introducing our canonical quadratic differential
quantities~\eqref{quad_zeta}, we find agreement with our expression for
${}^{(2)}\!{\bdelta}_{\mathrm s}$, except that~\cite{matetal98} appears to
differ by an overall factor of $\sfrac12$. Further, as regards the Poisson
gauge we find complete agreement with~\cite{baretal05} and~\cite{hwaetal12},
but encounter discrepancies with numerical coefficients in~\cite{matetal98}.

\section{Discussion} \label{Sec:concl}

In this paper we have given simple expressions for the second order
fractional density perturbation ${}^{(2)}\!{\bdelta}_{\bullet}$ for flat FLRW
cosmologies containing dust and a cosmological constant in four gauges:
Poisson, uniform curvature, total matter and synchronous. The unified form of
our expressions is due to our novel way of representing the temporal and
spatial dependence. First, as regards temporal dependence the expressions
depend algebraically on the function $g(x)$ that determines the first order
perturbations and on one other function ${\mathcal B}(x)$ that is defined as
an integral involving $g(x)$ and the background variables (see
eq.~\eqref{cal_Bfinal}). Second, we have introduced a new way of representing
the spatial dependence of the perturbations using the set of quadratic
differential expressions given in~\eqref{quad_zeta}.

Our representation shows that ${}^{(2)}\!{\bdelta}_{\bullet}$ assumes its
simplest form in the total matter and synchronous gauges, with the
super-horizon part being zero and the terms in the post-Newtonian part having
a common time dependence of $xg(x)$. On the other hand, the form in the
Poisson gauge is the most complicated. For example, the function ${\mathcal
B}(x)$ appears in the post-Newtonian part as well as in the Newtonian part,
unlike in the other gauges.

The form of our expressions for ${}^{(2)}\!\bdelta_{\bullet}$ makes it easy
to understand their overall dynamics and to read off their asymptotic
behaviour near the initial singularity $(x\rightarrow 0)$ and at late times
($x\rightarrow \infty)$. It follows from~\eqref{fg} and~\eqref{cal_Bfinal}
that
\begin{equation}
\lim_{x\rightarrow 0} g(x)=\sfrac35, \qquad
\lim_{x\rightarrow 0} {\mathcal B}(x)=\sfrac{1}{21}. \label{lim_0}
\end{equation}
In addition if $\Lambda>0$ then~\eqref{calH} yields ${\cal H}=O(x)$ and
$\Omega_m=O(x^{-3})$ as $x\rightarrow\infty$, which implies that
\begin{equation}
\lim_{x\rightarrow \infty} xg(x) \quad \text{and}\quad
\lim_{x\rightarrow \infty} {\mathcal B}(x) \quad \text{are finite and non-zero}. \label{lim_infty}
\end{equation}
Using these results, and the definitions~\eqref{fg} and~\eqref{cal_Bfinal},
we conclude that $g(x),{\mathcal B}(x)$ and $g^{-1}\Omega_m$ are positive and
bounded for $0<x<\infty$, and hence that the coefficients $A_{i,{\bullet}}$,
as given by~\eqref{Ai_p}-\eqref{Ai_v}, are bounded. It follows
from~\eqref{delta_2} and~\eqref{deltaN_2s} that the overall evolution of
${}^{(2)}\!\bdelta_{\bullet}$ is governed by the factors $xg$ and $x^2g^2$ in
the post-Newtonian and Newtonian terms.

As regards asymptotics, equations~\eqref{Eds_3} and~\eqref{lim_0} imply that
near the initial singularity ${}^{(2)}\!\bdelta_{\bullet}$ is approximated by the
Einstein-de Sitter specialization~\eqref{delta_2_eds}-\eqref{delta_2s_eds}.
It follows from~\eqref{delta_2_eds} and~\eqref{delta_2s_eds} that if
$\Lambda=0$ then ${}^{(2)}\!\bdelta_{\bullet}$ is unbounded at late times:
${}^{(2)}\!\bdelta_{\bullet} = O(x^2)$ as $x\rightarrow \infty$, with the Newtonian
part being dominant. In conjunction with the linear behaviour,
${}^{(1)}\!\bdelta_{\bullet} = O(x)$ as $x\rightarrow \infty$,
this suggests that the Einstein-de Sitter universe is unstable to scalar
perturbations. On the other hand, if $\Lambda>0$ eq.~\eqref{lim_infty}, in
conjunction with~\eqref{delta_2}-\eqref{deltaN_2s}, implies that both
${}^{(1)}\!\bdelta_{\bullet}$ and ${}^{(2)}\!\bdelta_{\bullet}$ are finite for all four
gauges when $x\rightarrow \infty$. This suggests that the \emph{presence of a
positive cosmological constant stabilizes the perturbations}, and that this
is the case for any of the four gauges under consideration. We can thus say
that the conclusion reached in~\cite{uggwai13} (see equations (2) and (3))
concerning stabilization is \emph{gauge-robust}. However, the choice of gauge
does affect the limit of ${}^{(2)}\!\bdelta_{\bullet}$ as $x\rightarrow \infty$.
Specifically it follows from~\eqref{Ai_p}-\eqref{deltaN_2s} and the
above-mentioned asymptotic properties of $g$ and $\Omega_m$ that
\begin{equation}
\lim_{x\rightarrow \infty}({ A}_1,{A}_2,{ A}_3,{ A}_4,{ A}_5)_{\bullet}=
\left(3(1+2a_{\mathrm{nl}}), 0, 0, -2, 1-2a_{\mathrm{nl}} \right),
\end{equation}
for  the Poisson and uniform curvature gauges and
\begin{equation}
\lim_{x\rightarrow \infty}({ A}_1,{A}_2,{ A}_3,{ A}_4,{ A}_5)_{\bullet}=
\left(0, 0, -5, 0, 2(2-a_{\mathrm{nl}}) \right),
\end{equation}
for the total matter and synchronous gauges. Observe that the effect of the
non-Gaussianity, as represented by $a_{\mathrm{nl}}$, persists indefinitely
into the future through the post-Newtonian term and in the first case also
through the super-horizon term.

In the course of doing the research reported in this paper we made a
detailed comparison of our results with the known expressions for
${}^{(2)}\!\bdelta_{\bullet}$ in the synchronous and Poisson gauges when the
background spatial curvature is zero. Our new canonical representation of the
spatial dependence has enabled us to unify, simplify and extend seemingly
disparate results, while at the same time revealing a number of errors in the
literature. One example concerns the Poisson gauge. Two expressions for
${}^{(2)}\!\bdelta_{\mathrm p}$ with $\Lambda>0$ have been given, which
appear to be completely different. The first, by Tomita~\cite{tom05}, was
derived by solving the perturbation equations in the synchronous gauge and
then transforming to the Poisson gauge. The second, by Bartolo \emph{et
al}~\cite{baretal10}, was derived by solving the perturbation equations
directly in the Poisson gauge. However, by simplifying the ${B}$-functions of
Bartolo and introducing our canonical representation of the spatial
dependence we have been able to show, after correcting some typos, that both
of these expressions can be written in our form for
${}^{(2)}\!{\bdelta}_{\mathrm p}$, which is given by~\eqref{delta_2}
and~\eqref{Ai_p}.

Another example concerns the synchronous gauge. Three expressions for
${}^{(2)}\!\bdelta_{\mathrm s}$ with $\Lambda>0$ have been
given~\cite{tom05,baretal10,bruetal13}, which are not related in an obvious
way. We have shown that the expression given by Tomita~\cite{tom05} can be
written in our form, given by~\eqref{deltaN_2s}.  As regards~\cite{baretal10}
and~\cite{bruetal13} the post-Newtonian part can be written in our form, but
in~\cite{baretal10} the Newtonian term is an approximation, while
in~\cite{bruetal13} the temporal and spatial dependence is not given in fully
explicit form.

To the best of our knowledge an expression for ${}^{(2)}\!{\bdelta}_{\bullet}$ with
$\Lambda>0$ has not been given previously for gauges other than the Poisson
and  synchronous gauges. However, Hwang \emph{et al}~\cite{hwaetal12} have
given an expression for ${}^{(2)}\!{\bdelta}_{\bullet}$ with $\Lambda=0$ using the
total matter and uniform curvature gauges (see equations (43) and (53)).
After converting the spatial dependence to our canonical form we obtain
agreement with our expressions. Details about all the above comparisons are
given in~\cite{uggwai13c}.

\subsection*{Acknowledgements}
We thank Marco Bruni and David Wands for helpful correspondence concerning
their recent paper~\cite{bruetal13}.  CU also thanks the Department of
Applied Mathematics at the University of Waterloo for kind hospitality. JW
acknowledges financial support from the University of Waterloo.


\begin{thebibliography}{99}


\bibitem{baretal04b} N.~Bartolo, E.~Komatsu, S.~Matarrese and A.~Riotto.
\newblock Non-Gaussianity from inflation: theory and observations,
\newblock  Physics\ Reports\ {\bf 402}, 103-266 (2004).

\bibitem{baretal10b}	N.~Bartolo, S.~Matarrese and A.~Riotto.
\newblock Non-Gaussianity and the Cosmic Microwave Background Anisotropies.
\newblock  Advances in Astronomy, {\bf 2010}, 157079, (2010), arXiv:1001.3957 [astro-ph.CO].

\bibitem{pitetal10} C.~Pitrou, J-P.~Uzan,  and F.~Bernardeau.
\newblock The cosmic microwave background bispectrum from
the non-linear evolution of the cosmological perturbations.
\newblock JCAP\ {\bf1007}, 003 (2010).

\bibitem{matetal98} S.~Matarrese, S.~Mollerach and M.~Bruni.
\newblock Relativistic second-order perturbations of the Einstein-de Sitter
universe.
\newblock Phys.\ Rev.\ D {\bf 58}, 043504 (1998).

\bibitem{baretal05} N.~Bartolo, S.~Matarrese and A.~Riotto.
\newblock Signatures of Primordial Non-Gaussianity in the Large-Scale Structure of
the Universe.
\newblock JCAP\ {\bf 0510}, 010 (2005).

\bibitem{tom05} K.~Tomita.
\newblock Relativistic second-order perturbations of
nonzero-$\Lambda$ flat cosmological models and CMB anisotropies.
\newblock Phys.\ Rev.\ D\ {\bf 71}, 083504 (2005).

\bibitem{baretal06}	N.~Bartolo, S.~Matarrese and A.~Riotto.
\newblock The full second-order radiation transfer function for large-scale CMB
anisotropies.
\newblock JCAP\ {\bf 0605}, 010 (2006).

\bibitem{baretal10} N.~Bartolo, S.~Matarrese, O.~Pantano and A.~Riotto.	
\newblock Second-order matter perturbations in a $\Lambda$CDM cosmology and
non-Gaussianity.
\newblock Class.\ Quantum\ Grav.\ {\bf 27}, 124009 (2010).


\bibitem{bruetal13} M.~Bruni, J.~C.~Hidalgo, N.~Meures and D.~Wands.
\newblock Non-Gaussian initial conditions in $\Lambda$CDM: Newtonian,
relativistic and primoridial contributions.
\newblock arXiv:1307.1478v1.

\bibitem{uggwai13} C.~Uggla and J.~Wainwright.
\newblock Asymptotic analysis of perturbed dust cosmologies to second order.
\newblock Gen.\ Rel.\ Grav.\ {\bf 45} 1467 (2013) doi:10.1007/s10714-013-1559-0.

\bibitem{uggwai13c} C.~Uggla and J.~Wainwright.
\newblock Second order density perturbations for dust cosmologies.
\newblock  arXiv:0910136

\bibitem{uggwai12} C.~Uggla and J.~Wainwright.
\newblock Scalar Cosmological Perturbations.
\newblock Class.\ Quantum\ Grav.\ {\bf 29} 105002 (2012) doi:10.1088/0264-9381/29/10/105002.

\bibitem{malwan09} K.~A.~Malik and D.~Wands.
\newblock Cosmological perturbations.
\newblock Physics\ Reports\ {\bf 475}, 1-51 (2009).

\bibitem{tom67} K.~Tomita.
\newblock Non-linear theory of gravitational instability in an expanding universe.
\newblock Prog.\ Theor.\ Phys.\ {\bf 37}, 831 (1967).

\bibitem{hwaetal12} J-C.~Hwang, H.~Noh, and  J-O.~Gong.
\newblock Second order solutions of cosmological perturbation in the matter dominated era.
\newblock Astrophys.\ J.\ {\bf 752}, 50 (2012) doi:10.1088/0004-637X/752/1/50.

\end{thebibliography}
\end{document}